\documentclass[11pt,twoside]{article} 
\usepackage{asp2004}
\usepackage{epsf}
\usepackage{psfig}
\usepackage{lscape}

\begin{document} 
\title{On Possible Oxygen/Neon White Dwarfs: H1504+65 and the White Dwarf Donors
  in Ultracompact X-ray Binaries}

 \author{K. Werner,$^1$ N.J. Hammer,$^1$ T. Nagel,$^1$ T. Rauch,$^{1,2}$ and S. Dreizler$^3$}
 \affil{$^1$Institut f\"ur Astronomie und Astrophysik, Universit\"at T\"ubingen,
 Sand~1, 72076 T\"ubingen, Germany\\
 $^2$Dr.~Remeis-Sternwarte, Sternwartstr.~7, 96049 Bamberg, Germany\\
 $^3$Institut f\"ur Astrophysik, Universit\"at G\"ottingen,
  Friedrich-Hund-Platz~1, 37077 G\"ottingen, Germany}

\begin{abstract} 
We discuss the possibility to detect O/Ne white dwarfs by evidence for Ne
overabundances. The hottest known WD, H1504+65, could be the only single WD for
which we might be able to proof its O/Ne nature directly. Apart from this,
strong Ne abundances are known or suspected only from binary systems, namely
from a few novae, and from a handful of LMXBs.  We try to verify the hypothesis
that the latter might host strongly ablated O/Ne WD donors, by abundance
analyses of the accretion disks in these systems. In any case, to conclude on
O/Ne WDs just by strong Ne overabundances is problematic, because Ne enrichment
also occurs by settling of this species into the core of C/O WDs.
\end{abstract}

\section{Introduction}

Stellar evolution theory predicts the existence of massive O/Ne white dwarfs as
a result of carbon burning. WDs with masses exceeding 1\,M$_\odot$ should be
such O/Ne WDs, but there are other ways to produce massive WDs, namely by binary
merging or by mass accretion during close-binary evolution. The question is: Is
there evidence for the existence of O/Ne WDs? A very useful overview about this
topic has been given by Weidemann (2003) at the last WD workshop at Naples. We
refer to his paper for a detailed discussion and references concerning stellar
evolution scenarios, which we only briefly summarize here.

O/Ne WDs should have masses of 1--1.3\,M$_\odot$. Below a core mass of
1\,M$_\odot$ C burning cannot ignite. If the core mass exceeds 1.3\,M$_\odot$,
an electron capture collapse will occur and probably result in a neutron
star. There is a debate as to what initial-mass range this 1--1.3\,M$_\odot$
core-mass range corresponds. Depending on details in evolutionary models, the
range of initial masses is between 9--11\,M$_\odot$ or between 7--9\,M$_\odot$
(for solar metallicity). In any case, in order to stop the O/Ne-core to grow (by
C burning) beyond the 1.3\,M$_\odot$ limit, the star must loose mass, either by
wind mass loss or by envelope removal in close binary systems.  The WD mass
distribution does show the existence of such massive  WDs, however, their origin
is not clear. It is possible that they are binary mergers.

If O/Ne WDs exist, then they have C/O envelopes and, thus, they are
spectroscopically indiscernible from C/O WDs. Possible evidence, however, are
observed neon overabundances in several cases: in neon novae, in some low-mass
X-ray binaries, and in one individual object (H1504+65). We will discuss here
the LMXBs and H1504+65. In the case of neon novae, only those very few objects
with extreme neon overabundances (over $\approx$30 times solar, i.e.\ 3\% mass
fraction) suggest O/Ne WDs, because up to 3\% $^{22}$Ne is expected in C/O WDs.

For the interpretation of Ne overabundances in novae or LMXBs it is important to
consider gravitational settling of Ne into the core of a C/O WD. This problem is
not yet solved theoretically. It could well be, that Ne overabundances observed
in eroded WD cores are the result of this process and not the result of C
burning.
   
\section{The hottest known white dwarf: H1504+65}

Spectroscopically, H1504+65 is an extreme member among the PG1159 stars, which
are hot, H-deficient objects on or closely before the hot end of the WD cooling
sequence. They are probably the result of a late He-shell flash that causes
ingestion and burning of H, and the exposition of He/C/O-rich intershell matter
on the surface (e.g.\ Werner 2001). H1504+65 is an extreme, and unique, PG1159
star because it is also He-deficient. The first quantitative analysis revealed
that the photosphere is mainly composed of C and O, by equal parts (Werner
1991), and it has been discussed that H1504+65 is either a naked C/O core or
that we see the C/O envelope of a O/Ne WD. The latter possibility is supported
by the fact that H1504+65 is the most massive known PG1159 star, although the
spectroscopic mass determination suggests a mass slightly below
1\,M$_\odot$. From the discovery of Ne lines in optical and EUV spectra (Werner
\& Wolff 1999) we found Ne=2--5\%.  This result is inconclusive concerning the
nature of the star as a possible O/Ne WD.

Soft X-ray and FUV spectra of H1504+65 were taken by Chandra/LETG and FUSE,
respectively, and the following atmospheric and stellar parameters were derived
(Werner et al.\ 2004):
$$
\mbox{$T_\mathrm{eff}$} = 200\,000\,{\rm K} \pm 20\,000\,{\rm K} 
\qquad \mbox{$\log g$} = 8.0 \pm 0.5 {\rm \ \ [cgs]}
$$
$$
{\rm C}=48 \quad
{\rm O}=48 \quad
{\rm Ne}=2 \quad
{\rm Mg}=2 \quad
{\rm He}<1 \quad
{\rm Na}<0.1 \quad
{\rm Al}<0.1
$$
$$
M/{\rm M}_\odot=0.836^{+0.13}_{-0.10} \qquad
\log L/{\rm L}_\odot=2.45^{+0.6}_{-0.4} \qquad
d/{\rm kpc}=0.67^{+0.3}_{-0.53}
$$
\noindent
Estimated errors for abundances are: $\pm$20\% for mass fraction of C and O, and
a factor of 3 for Ne and Mg. One significant result is the detection of Mg in
high amounts, which seems to confirm the O/Ne WD nature of H1504+65. However, a
detailed discussion reveals that the abundance determinations of Ne and Mg still
cannot decide if H1504+65 exposes a C/O core or a C/O envelope. The final answer
to this problem was hoped to be accomplished by UV spectroscopy with HST.
Planned observations during Cycle~13 aimed at the detection of Na which, if
strongly overabundant, is a clear indicator for C burning. Unfortunately, the
STIS spectrograph aboard HST died before the observations were carried out. It
appears that this question will remain unanswered for many years to come. This
is disappointing, because H1504+65 is the only object for which we can hope to
proof that single O/Ne WDs do really exist.

Considering our mass determination, even if we stretch the result to the limit
of our error bar, the mass of H1504+65 is slightly below 1\,M$_\odot$. However,
we point out that the evolutionary tracks which we used to derive the mass, are
those of post-AGB remnants that have lastly burned helium but -- H1504+65 has no
helium. There are no appropriate evolutionary tracks available. The main problem
is that we have no idea what is the evolutionary history of this unique star:
What event caused the helium-deficiency? We finally note that there is no
indication that H1504+65 is in a binary system.

\section{WDs in ultracompact low-mass X-ray binaries}

It could be that AGB-like mass loss of single stars during carbon burning is not
sufficient to stop the growth of the O/Ne core before the onset of
electron-capture collapse. Hence, close binary systems in which mass is
transferred from the C-burning star onto a companion might be the only places
where O/Ne WDs are created. First evidence for the existence of such systems
came from neon overabundances detected by line emission in (low resolution) ASCA
spectra of the LMXB 4U\,1626-67 (Angelini et al.\ 1995). Evidence for neon-rich
absorbing circumstellar material was found in Chandra LETG spectra of
4U\,0614+091 and three other LMXBs (Juett et al.\ 2001). The detection of
double-lined emissions in the Chandra HETG spectrum of 4U\,1626-67 (Schulz et
al.\ 2001) suggests neon and oxygen overabundances in the accretion disk of this
system.

In total there is now a group of five LMXBs, for which it is believed that they
are ultracompact systems (orbital periods below 1 hour, i.e.\ separation of the
order 1 light-s) in which a neutron star or a black hole accretes matter from a
low-mass ($\approx 0.1$\,M$_\odot$) degenerate star, the strongly ablated core
of a WD. Mass transfer is driven by gravitational radiation.

The luminosity of these systems is dominated by the accretion disk. In the soft
X-ray and UV spectral regions we see the hot (several 100\,000\,K) innermost
parts of the disk, which can be X-ray heated by the neutron star. The optical
light is dominated by cooler (about 10\,000\,K), outer parts of the disk. So we
have the unique possibility to study the former WD interior composition by the
determination of element abundances in the accretion disk.

For this purpose we have obtained optical medium-resolution spectra of two of
these systems (4U\,1626-67 and 4U\,0614+091). We have used the FORS1
spectrograph attached to UT1 of ESO's Very Large Telescope to record long-slit
spectra. We used two grisms (600B and 600R) and obtained spectra in the regions
3450--5900\AA\ and 5250--7450\AA\ with a resolution of about 5\AA. The B
magnitude of both targets is near 18.7; exposure times for both binaries were 87
and 58 minutes for the red and the blue spectra, respectively. The observations
were performed in service mode during Nov.~2003 and Mar.~2004. The spectra were
reduced using standard IRAF procedures. They are presented in Fig.\,1, and show
emission lines from C\,II/III and O\,II/III. They lack hydrogen and helium
lines, apparently confirming the H- and He-deficiency of the donor stars. In the
case of 4U\,0614+091 we confirm the line identifications of Nelemans et
al. (2004), who have obtained VLT/FORS2 spectra with slightly higher resolution.

In addition to these data, we also make use of archival HST/UV data of
4U\,1626-67, kindly provided in reduced form by Homer et al.\ (2002). The HST
spectrum is shown in Fig.\,2. It shows emission lines from C\,IV, O\,V, and
Si\,IV. It does not show He\,II~1640\AA.

\begin{landscape}
\begin{figure}[!t]
\epsfysize=11cm \epsffile{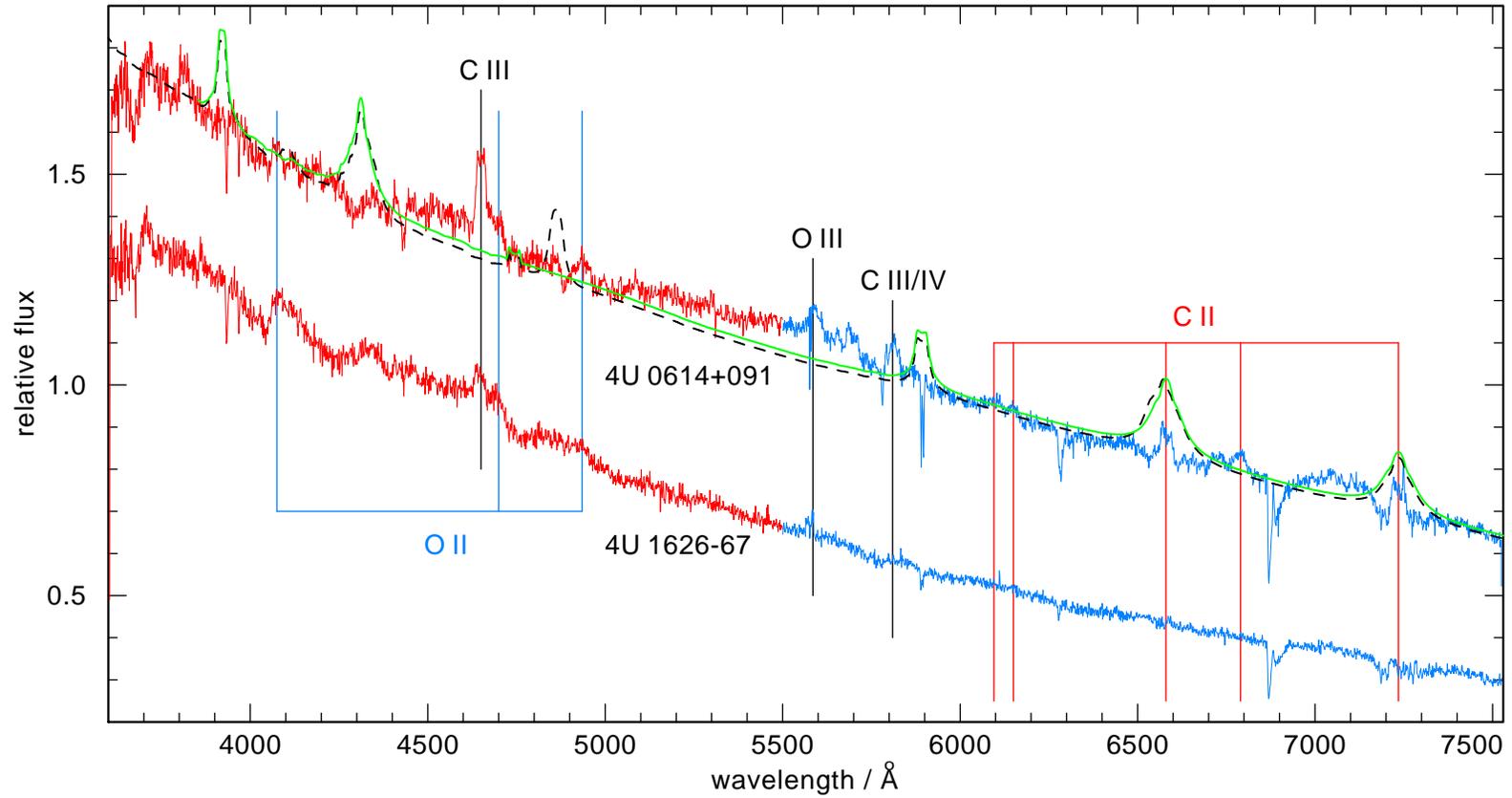}
\caption{VLT spectra of the probably H- and He-deficient disks in two ultracompact
  LMXBs. Some line features are identified. Overplotted are two models matching
  qualitatively some emission lines. One model (dashed line) includes trace
  hydrogen and shows a distinct H$\beta$ emission, which is not observed. The
  model emission lines at 3920\,\AA\ and 4310\,\AA, which are not seen in the
  observations, are from C\,II.}
\end{figure}
\end{landscape}

We have begun the construction of accretion disk models to calculate synthetic
spectra and report here on the current state of our work. We use our accretion
disk code AcDc, which is described in detail by Nagel et al.\ (2004). In
essence, it assumes a radial $\alpha$-disk structure (Shakura \& Sunyaev
1973). Then the disk is divided into concentric annuli. For each annulus we
solve the radiation transfer equation (assuming plane-parallel geometry)
together with the NLTE rate equations, plus energy- and hydrostatic equations,
in order to calculate a detailed vertical structure. The integrated disk
spectrum is then obtained by co-adding the spectra from the individual
annuli. In comparison with the observed UV/optical spectra we present here first
model spectra for particular, representative disk regions.

\begin{figure}[!t]
\epsfxsize=\textwidth \epsffile{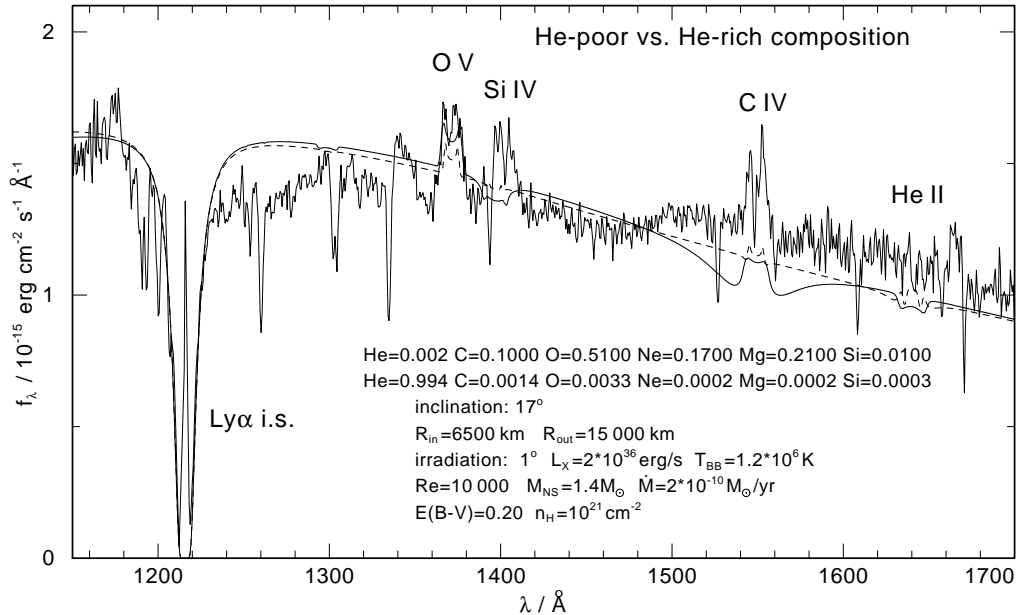}
\caption{HST/STIS accretion disk spectrum of 4U\,1626-67 and two model 
spectra, differing by their helium abundance. One is He-poor (0.2\% by mass,
full line), the other is He-rich (99\%, dashed line). The observed lack of
\ion{He}{II}~1640\AA\ obviously is no proof for He-deficiency of the disk. Note
the broad absorption wings of the \ion{C}{IV} resonance line in the He-poor
model, which are not observed. This points at a stronger irradiation than
assumed.}
\end{figure}

\subsection*{4U\,1626-67 and 4U\,0614+091}

4U\,1626-67 has an orbital period of 42~min (see e.g.\ Chakrabarty 1998, and
references therein). The accretor is a 7.7s X-ray pulsar with a magnetic field
strength of $3\cdot 10^{12}$\,G. The donor's mass is smaller than
0.1\,M$_\odot$. The system separation is 300\,000\,km. The inner disk edge is at
the co-rotation radius, at R=6500\,km from the pulsar, and the outer edge at the
tidal truncation radius, near R=200\,000\,km. The mass transfer rate is roughly
$2\cdot 10^{-10}$\,M$_\odot$/yr and the X-ray luminosity is about $2\cdot
10^{36}$\,erg/s.

Fig.\,2 shows the UV spectrum of two disk models extending from
R=6500--15\,000\,km, seen under an inclination angle of 17$\deg$. They are
irradiated by the neutron star, assuming the X-ray luminosity mentioned above
and a blackbody energy distribution (T$=1.2\cdot 10^6$\,K). The two models
differ by their chemical composition. One model is He-poor (dominated by C, O,
Ne, Mg), the other is He-rich with abundances typical for He-rich disks in
AM~CVn systems. This comparison shows that neither model exhibits a clearly
detectable He\,II~1640\,\AA\ line. Hence the lack of this line in the HST
spectrum is no proof of He-deficiency. The situation is perhaps more favorable
in the optical region. In Fig.\,1 we have plotted two model spectra from C/O
dominated disk regions at R=70\,000\,km, one without H and the other with trace
H ($10^{-6}$ by number). A strong H$\beta$ emission is seen in the second case,
which is not present in the observed spectra. Although we have not yet performed
this test with helium, we hope that we can derive a tight upper limit for the He
abundance from the lack of He\,I lines.

The parameters of the system 4U\,0614+091 are poorly known. No orbital period
was measured. From X-ray bursts it is assumed that the accretor is a neutron
star. From the similarity of the X-ray spectral characteristics with
4U\,1626-67, Juett et al.\ (2001) suggest that this system is also ultracompact.
The optical spectrum displays a number of C and O lines, as already identified
by Nelemans et al.\ (2004). The lack of He\,I and H\,I lines supports the
suggestion of Juett et al.\ (2001).  The two C/O dominated models shown in
Fig.\,1 were computed for this system. Irradiation by the central object is not
taken into account. They do show emission features, which qualitatively match
some observed features. It appears that the disk in this region is optically
thin.

Both LMXBs discussed here do not show neon lines in the optical
spectra. Detailed modeling has to be performed to conclude what this means for
the maximum possible neon abundance.

\acknowledgements{T.R.~is supported by DLR under grant 50\,OR\,0201.}

\end{document}